\documentclass{optica-article}
\journal{opticajournal}
\articletype{Research Article}

\usepackage{subcaption}

\usepackage{lineno}

\usepackage[linesnumbered,ruled,vlined]{algorithm2e}
\makeatletter
\providecommand{\doi}[1]{%
  \begingroup
    \let\bibinfo\@secondoftwo
    \urlstyle{rm}%
    \href{http://dx.doi.org/#1}{%
      doi:\discretionary{}{}{}%
      \nolinkurl{#1}%
    }%
  \endgroup
}
\makeatother
\usepackage[nameinlink,noabbrev,capitalize]{cleveref}

\renewcommand{\cref}{\Cref}

\let\oldexp\exp
\renewcommand{\exp}[1]{\oldexp{\left( {#1} \right)}}

\begin{document}
	\title{HoloTile RGB: Ultra-fast single-shot, discretized, full-color computer-generated phase-only holography}
	
	\author{\authormark{*}Andreas Erik Gejl Madsen, Jesper Gl\"uckstad}

	\address{SDU Centre for Photonics Engineering\\University of Southern Denmark\\DK-5230 Odense M, Denmark}
	
	\email{\authormark{*}gejl@mci.sdu.dk}
	
\begin{abstract*}
	We demonstrate the first use of the HoloTile Computer-Generated Holography (CGH) modality on multi-wavelength targets. Taking advantage of the sub-hologram tiling and Point Spread Function (PSF) shaping of HoloTile allows for reconstruction of high-fidelity, pseudo-digital multi-wavelength images, with well-defined discrete output pixels, without the need for temporal averaging. 
	For each wavelength, the target channels are scaled appropriately, using the same output pixel size. We employ a stochastic gradient descent (SGD) hologram generation algorithm for each wavelength, and display them sequentially on a HoloEye GAEA 2.1 Spatial Light Modulator (SLM) in Color Field Sequential (CFS) phase modulation mode. As such, we get full 8-bit phase modulation at 60 Hz for each wavelength. The reconstructions are projected onto a camera sensor where each RGB image is captured in a single shot.
	While these show impressive color reconstructions, the method can be adapted to any wavelength combination for use in a plethora of multi-wavelength application.
\end{abstract*}

\section{Introduction}
\noindent
The addition of multi-wavelength illumination to any field employing some version of structured or modulated light provides opportunities for greater control, flexibility, and performance.
Aside from allowing for color rendition using red-green-blue (RGB) illumination for display, projection, and AR/VR\footnote{Augmented/Virtual Reality} purposes, structured multi-wavelength illumination has found uses in advanced microscopy \cite{mico_single-shot_2023,li_multi-wavelength_2022}, particle trapping and manipulation \cite{wu_single-fiber_2022,freitag_flying_2024}, optogenetics and neuro-photonics \cite{kampasi_dual_2018,vierock_bipoles_2021,papagiakoumou_scanless_2010}, optical computing and communication \cite{li_massively_2023,kemal_multi-wavelength_2016}, and volumetric additive manufacturing (VAM) \cite{whyte_volumetric_2024,li_intensity-coupled_2024,de_beer_rapid_2019,alvarez-castano_holographic_2025,alvarez-castano_holographic_2024-1}, to name a few.

Computer-generated holography (CGH) presents several characteristics that directly benefit these applications.
While projection systems based on imaging amplitude modulating digital micromirror displays (DMDs) are commonplace and simple to implement, they are accompanied by several downsides, including limited light propagation and point-spread function (PSF) control, single-plane focus, and projection pattern dependent light efficiency.
In contrast, using a phase-only spatial light modulator (SLM) in tandem with a Fourier transforming lens system allows for nearly 100\% light efficiency in holographic reconstructions, independent of the target pattern. 
In addition, the phase control of the SLM and the holographic encoding allows for dynamical shaping of light in both two or three dimensions.

To fully leverage these benefits, we introduce multi-wavelength HoloTile, a CGH modality based on subhologram tiling and explicit PSF control. This approach is reconfigurable to any number of wavelengths, making it adaptable to the specific requirement of the aforementioned applications. While the wavelength selection can be freely tailored, in this paper, we focus on RGB holograms and their reconstruction due to their visual appeal.

HoloTile \cite{gluckstad_holographic_2022,madsen_holotile_2022,madsen_digital_2024,alvarez-castano_holographic_2025,madsen_efficient_2023,gluckstad_holotile_2024,madsen_axial_2025,gluckstad_holographic_2023} overcomes several key challenges of conventional and color CGH. Conventional CGH methods often suffer from severe speckle-noise due to the coherent illumination and limited consideration of the optical system PSF.
Moreover, the calculation of high-resolution holograms necessary to take advantage of modern SLMs is typically too slow to be calculated in any real-time manner.

The subhologram tiling and PSF shaping of HoloTile presents several advantages; first, holograms are exponentially faster to calculate, and second, the tiling causes the reconstruction to only be defined in a discrete point grid, effectively separating the spatial frequency components.
By adding a second, PSF shaping, hologram to the tiled hologram allows for complete control of the shape of each point in the reconstruction grid. By shaping the PSF to be square, the reconstructions can effectively be discretized into a pseudo-digital output pixel grid.

The well-defined output pixels do not overlap in the reconstruction plane and do therefore not contribute to the distinct speckle noise recognizable in conventional CGH.
As opposed to other implementations of CGH \cite{pi_review_2022}, and particular color CGH, it is not necessary to temporally average several holograms \cite{liu_speckle_2019,mori_speckle_2014,hsu_speckle_2011,takaki_speckle-free_2011,madsen_-axis_2023} in order to achieve homogeneous color reconstructions. This single-shot capability is key for dynamic and real-time applications, as it eliminates the need for multiple exposures, enabling faster and more efficient holographic generation and reconstruction.
Furthermore, since the sub-holograms are much smaller than the SLM resolution, calculation of HoloTile holograms can be performed at rates higher than $60$ Hz, allowing for real time calculation and display of scenes.

The pseudo-digital pixel-based formulation of HoloTile not only enables high-speed, high-quality holographic multi-wavelength reconstructions but also provides a fully scalable approach adaptable to any combination of wavelengths, ensuring broad applicability across diverse multi-wavelength applications.

\begin{figure}[h]
	\centering
	\includegraphics[width=\textwidth]{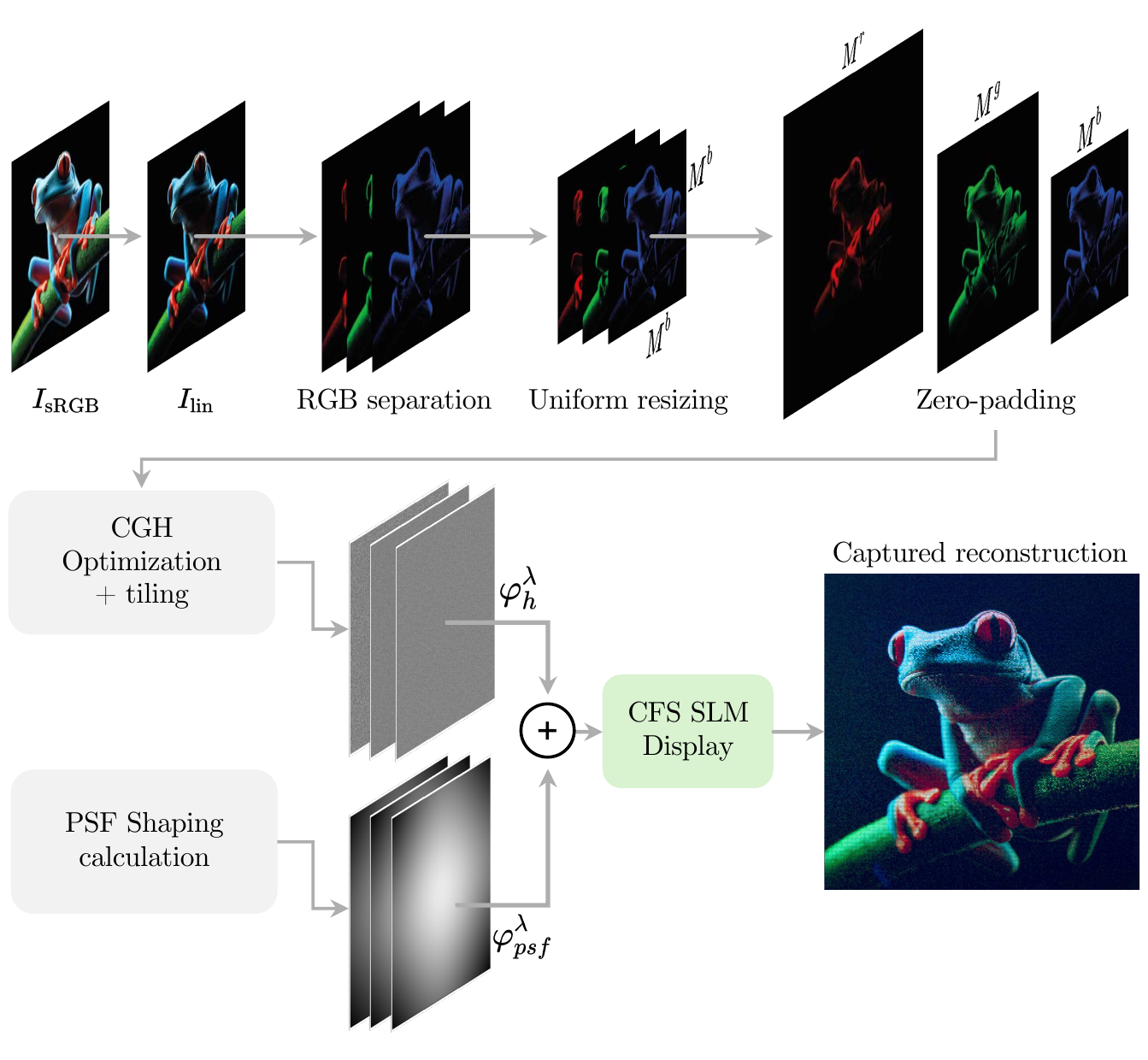}
	\captionsetup{width=.9\linewidth}
	\caption{Graphical overview of the HoloTile RGB process. An input RGB image is linearized, followed by processing of each individual color channel to facilitate co-localization of HoloTile output pixels. The calculated object holograms are combined with associated PSF shaping holograms and shown in rapid succession on a CFS SLM, after which the final reconstruction is captured in camera.}
	\label{fig:rgb-process}
\end{figure}

\section{Theory}
The process of generating HoloTile-based RGB computer-generated holograms follows a structured sequence of computational steps, ensuring both light efficiency and high-quality reconstructions. In this section, a chronological overview of the procedure is given (and illustrated in \cref{fig:rgb-process}), detailing the ordered calculations required for each individual multi-wavelength hologram generated for this paper. 

\subsection{Linear Color Space Conversion}
When the target image is loaded into memory, it is encoded in the sRGB color space. As a result, the pixel values are not linearly proportional to light intensity due to the non-linear gamma encoding. A captured reconstruction of an uncorrected sRGB target results in perceptual color and contrast distortions.
To ensure accurate intensity representation of the captured reconstructions, a linear color space intensity version of the same image is calculated.
The conversion from sRGB to linear color space is given by \cite{lee_high-contrast_2022}:
\begin{align}
I_{\text{lin}} = 
	\begin{cases} 
		\frac{I_{\text{sRGB}}}{12.92} & \text{for}  \quad 0 \leq I_{\text{sRGB}} \leq 0.04045 \\
		\left(\frac{I_{\text{sRGB}} + 0.055}{1.055}\right)^{2.4} & \text{for} \quad 0.04045 < I_{\text{sRGB}} \leq 1
	\end{cases}
\end{align}
Thus converting the sRGB intensity image to a linear space intensity for correct color and contrast representation.

\subsection{Output Grid Sampling and Alignment}
\begin{figure}[h]
	\centering
	\includegraphics{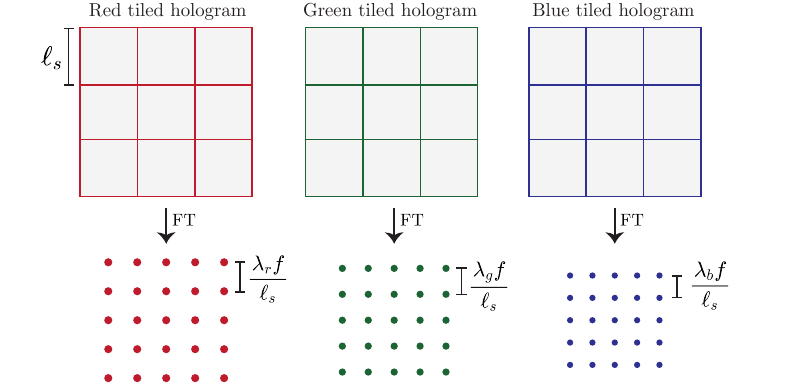}
	\captionsetup{width=.9\linewidth}
	\caption{Effect on output grid of identical subhologram sizes, regardless of illuminating wavelength. As the wavelength decreases, so does the grid spacing. FT denotes a spatial 2D Fourier transform.}
	\label{fig:rgb-hologram-spacing}
\end{figure}

\begin{figure}[h]
	\includegraphics{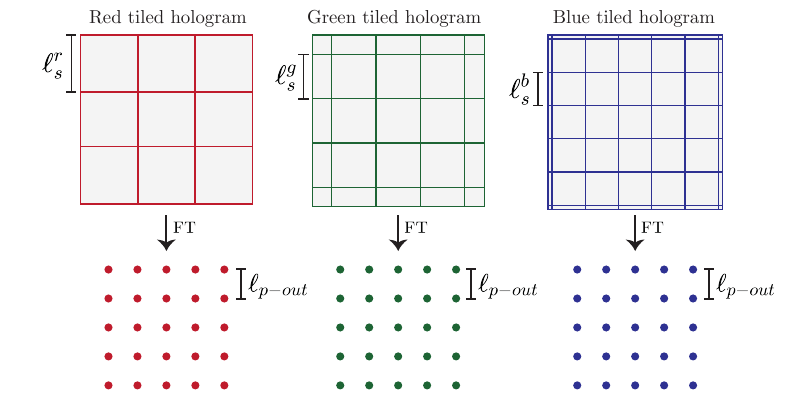}
	\captionsetup{width=.9\linewidth}
	\caption{Effect on output grid of wavelength-dependent subhologram sizes. As the illuminating wavelength decreases, so should the size of the subholograms, facilitating co-localization of the grid points. FT denotes a spatial 2D Fourier transform.}
	\label{fig:rgb-hologram-spacing-fixed}
\end{figure}

\noindent
The HoloTile modality defines an output grid in the reconstruction plane, with which the PSF shape is convolved. 
The output ``pixel" grid arises from the tiling of subholograms on the SLM. This grid can be expressed as
\begin{equation}
	A_f(x,y:t) \propto  A_\textrm{PSF}(x, y, z; t) \circledast  A_\textrm{sub}\left(x - \frac{m\lambda f}{\ell_s}, y - \frac{n\lambda f}{\ell_s}; t\right) \label{eq:conv}
\end{equation}
where $\circledast$ denotes spatial convolution, $\ell_s$ is the physical size of the subholograms, $\lambda$ is the illuminating wavelength, $f$ is the focal length of the Fourier transforming lens, $A_\textrm{PSF}$ is the Fourier transform of the PSF shaping hologram, and $A_\textrm{sub}$ is the Fourier transform of the subhologram. $m$ and $n$ are output grid indices occurring due to the tiling operation.
The reconstruction of the subhologram is therefore discretely sampled and thus only defined in specific locations defined by the spatial frequency comb array $\sum\limits_{m,n = -\infty}^{\infty} \delta\left(x - \frac{m\lambda f}{\ell_s}, y - \frac{n \lambda f}{\ell_s}\right)$.
However, the wavelength dependence of the sampling points in the output grid results in different grid spacings in the reconstruction, as is illustrated in \cref{fig:rgb-hologram-spacing}. This effectively corresponds to a color display with different pixel pitches for the three color channels. 
To make sure that the red, green, and blue channel of the reconstructions overlap exactly in the reconstruction plane -- maintaining identical pixel pitch -- the output pixel spacing must be the same for all three wavelengths:
\begin{equation}
	\frac{\lambda^r f}{\ell_s^r} = 	\frac{\lambda^g f}{\ell_s^g} = \frac{\lambda^b f}{\ell_s^b} = \ell_{p-out}
\end{equation}
This condition requires the resizing of the subholograms depending on their illuminating wavelength. The process begins by defining a desired pixel spacing, $\ell_{p-out}$, which serves as the  spacing for all three color channels. Given the target spacing, the physical sizes of the subholograms, and their corresponding resolutions, $M^\lambda$, can be determined by the following relations:
\begin{align}
\begin{alignedat}{2}
    \ell^r_s &= \frac{\lambda^r f}{\ell_{p-\text{out}}} &\quad \Rightarrow\quad & M^r = \left\lfloor \frac{\ell^r_s}{\ell_p} \right\rfloor \\
    \ell^g_s &= \frac{\lambda^g}{\lambda^r} \ell^r_s &\quad \Rightarrow\quad & M^g = \left\lfloor \frac{\ell^g_s}{\ell_p} \right\rfloor \\
    \ell^b_s &= \frac{\lambda^b}{\lambda^r} \ell^r_s &\quad \Rightarrow\quad & M^b = \left\lfloor \frac{\ell^b_s}{\ell_p} \right\rfloor
\end{alignedat}
\end{align}
Where $\ell_p$ is the pixel pitch of the SLM and $\lfloor\cdot\rfloor$ denotes the flooring operation. This will ensure the correct spacing of output pixels in the reconstruction plane for all three illuminating wavelengths, as illustrated in \cref{fig:rgb-hologram-spacing-fixed}. 

\begin{figure}[ht]
	\begin{subfigure}{\textwidth}
	\centering
	\includegraphics[width=\textwidth]{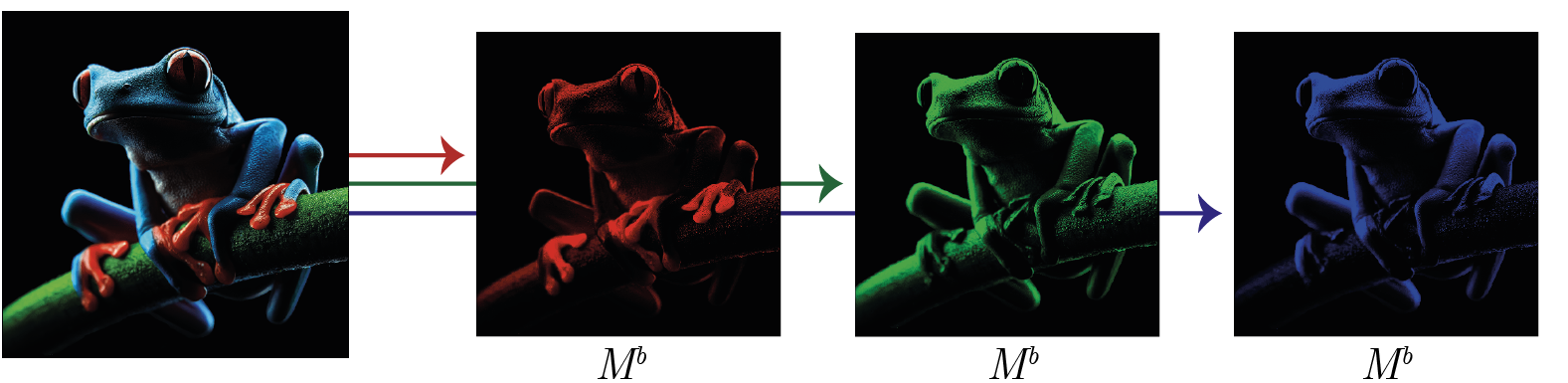}
	\caption{}
	\label{fig:rgb-color-resizing}
	\end{subfigure}
	\par\bigskip
	\begin{subfigure}{\textwidth}
	\centering
	\includegraphics[width=\textwidth]{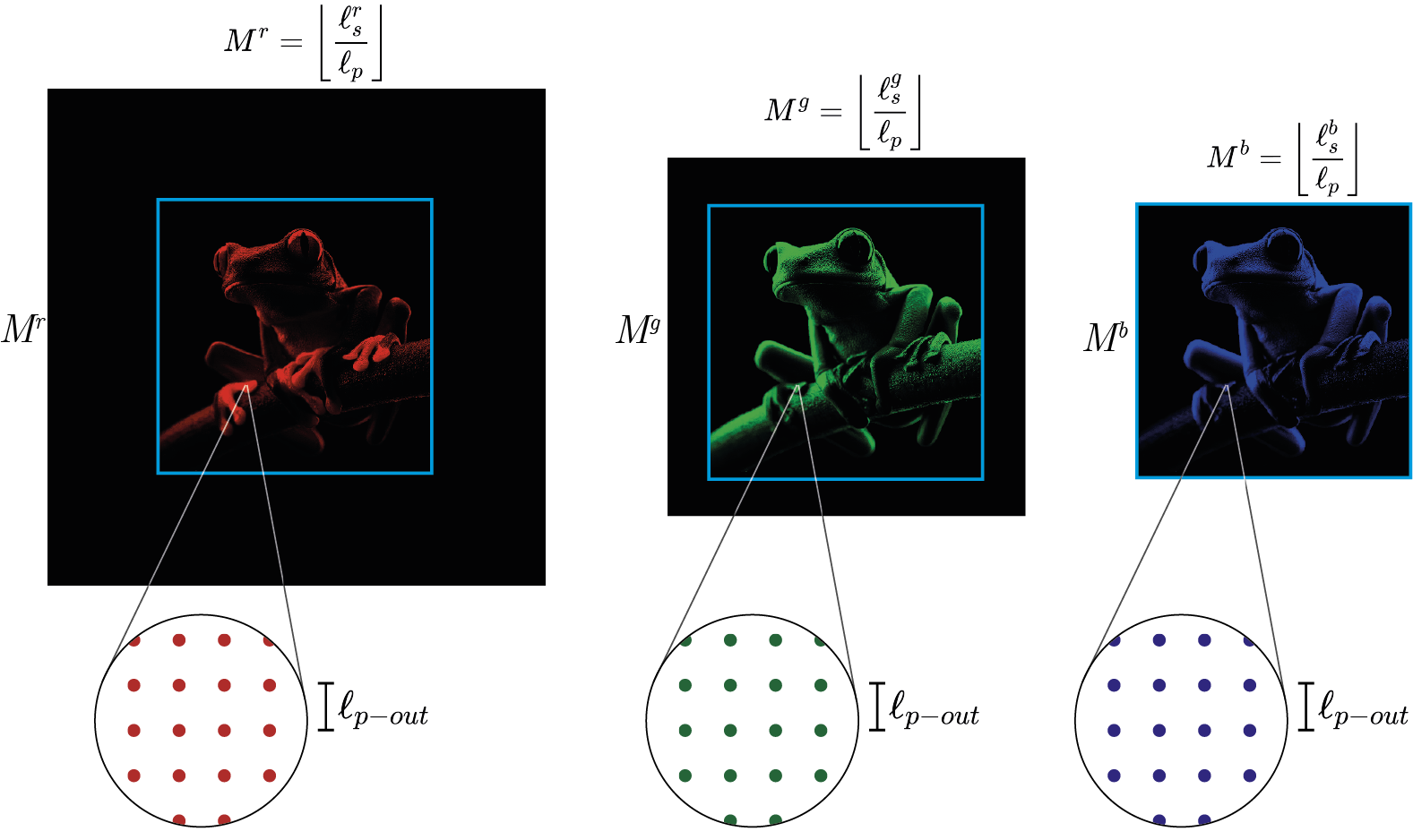}
	\caption{}
	\label{fig:rgb-color-spacing}
	\end{subfigure}
	\captionsetup{width=.9\linewidth}
	\caption{Illustration of uniform resizing and zero-padding of the color channels to ensure target co-localization. (a) The color channels are extracted and resized to the smallest dimensions, e.g., the dimensions of the blue subholograms, $M_b$. (b) Each color channel is zero-padded to its associated wavelength-dependent subhologram size, e.g., $M_r$, $M_g$, and $M_b$.}
	\label{fig:color-separation-alignment}
\end{figure}

However, simply resizing the color channels of the target image, $I_\text{lin}$, to these resolutions will, necessarily, cause scaling and alignment issues between the reconstructed color channels, as the output resolution will be different between the channels. To ensure proper co-localization in the final RGB reconstruction, a two-step adjustment (illustrated in \cref{fig:color-separation-alignment}) is required:
\begin{enumerate}
	\item \textbf{Uniform resizing} All three color channels in the target are resized to the common resolution of the smallest subhologram, $M^b$, ensuring that they share the same pixel grid.
	\item \textbf{Zero-padding} Each resized channel is then zero-padded to its respective resolution, i.e., $M^r$, $M^g$, and $M^b$. This zero-padding ensures that the \textit{subject}, the target image, is identical in size and resolution in the reconstruction.
\end{enumerate}
As a result, the final targets for hologram optimization consists of the three separate color channels, $I^r$, $I^g$, and $I^b$, with resolutions $M^r\times M^r$, $M^g\times M^g$, and $M^b\times M^b$, respectively.

\subsection{CGH Optimization}
\noindent
To synthesize the sequential RGB hologram, each pre-processed color channel target is used to generate an associated phase-only subhologram.
As HoloTile is agnostic towards the specific algorithm used for hologram generation, all reconstructions shown in this paper are calculated using stochastic gradient descent (SGD) as presented in \cref{alg:SGD}. 
However, as shown in previous publications \cite{madsen_holotile_2022,gluckstad_holotile_2024}, machine learning approaches or Gerchberg-Saxton-based (GS) \cite{gerchberg_practical_1972} algorithms are equally valid.
The calculated wavelength-specific subhologram phases, $\varphi_{sub}^\lambda$, are then tiled to fit the SLM resolution to create three tiled object holograms, $\varphi_h^\lambda$.

Reconstructing the tiled object holograms at this stage would result in reconstructions defined in points on the output grid due to the tiling operation, in the shape of the target. In order to create the pseudo-pixels in the output, PSF shaping is necessary.

\begin{algorithm}
\caption{SGD CGH Optimization}
\label{alg:SGD}
\SetKwInOut{Input}{Input}
\SetKwInOut{Output}{Output}

\Input{Target amplitude $A_t$, iterations $T$, learning rate $lr = 1e6$, parameter $\alpha = 10$.}
\Output{Optimized hologram phase $\varphi_h$.}

$\varphi_o \gets \text{random values}$ \textcolor{cyan}{// Initialize object phase} \\
$H = \text{FFT}(A_t \cdot e^{i\varphi_o})$ \textcolor{cyan}{// Compute initial hologram} \\
$\varphi_h = \angle H$ \textcolor{cyan}{// Extract hologram phase} \\
\For{$i = 1$ \KwTo $T$}{
    $E_r \leftarrow \text{IFFT}(e^{i\varphi_h})$ \textcolor{cyan}{// Propagate back to object} \\
    $A_r \leftarrow |E_r|$ \textcolor{cyan}{// Compute amplitude} \\
    $A_r \leftarrow A_r \cdot \sqrt{\frac{\sum A_t^2}{\sum A_r^2}}$ \textcolor{cyan}{// Correct the energy} \\
    $\mathcal{L} \leftarrow \text{MSE}(A_r^2, A_t^2)$ \textcolor{cyan}{// Calculate MSE loss} \\
    
    $\varphi_h \leftarrow \varphi_h - \text{lr} \cdot \nabla_{\varphi_h} \mathcal{L}$ \textcolor{cyan}{// Perform SGD update} \\
}
\Return{$\varphi_h$}
\end{algorithm}

\subsection{RGB PSF Shaping}
\begin{figure}[h]
	\centering
	\includegraphics[width=\textwidth]{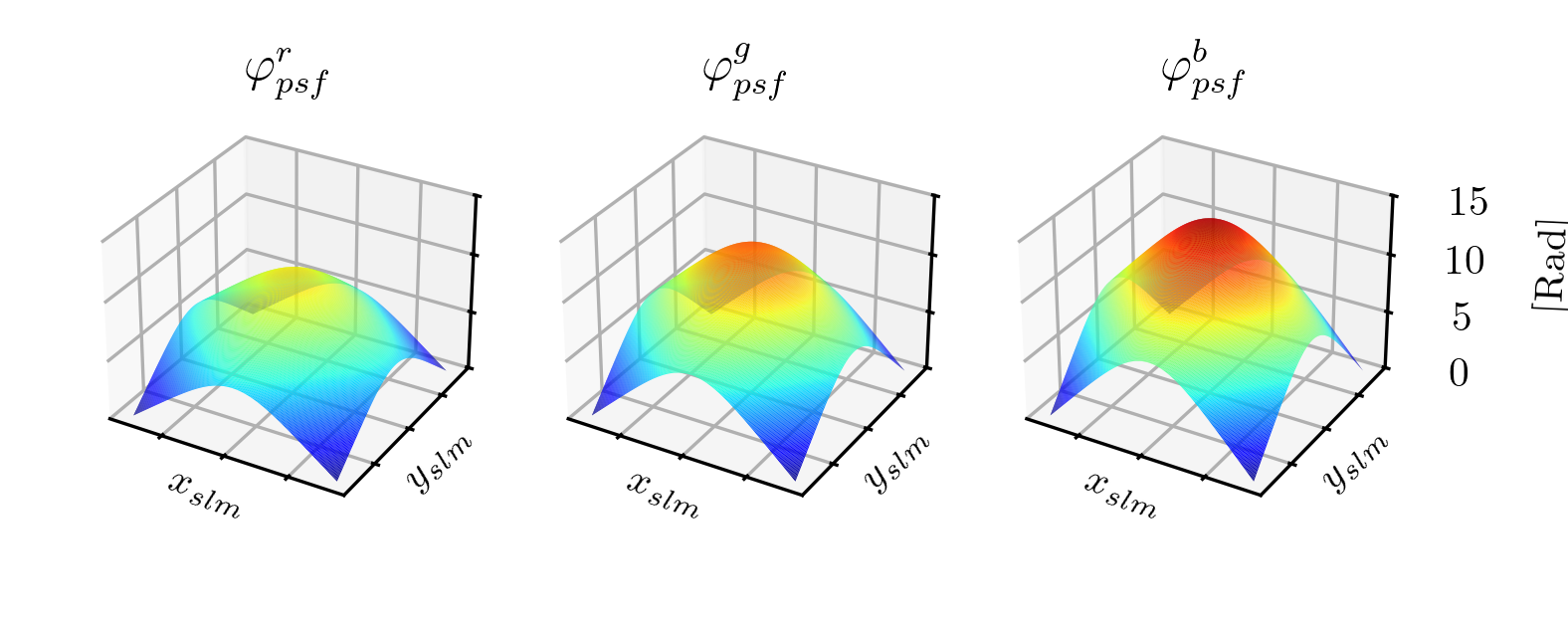}
	\captionsetup{width=.9\linewidth}
	\caption{Calculated PSF shaping phase profiles for the three illuminating wavelengths, all resulting in $\ell_{p-out} = 30\text{ }\mu m$ square output pixel sizes. }
	\label{fig:psf-shaping-rgb}
\end{figure}
To fill out the spatial frequency comb array that results from reconstructing only the tiled object holograms and create distinct output pixels, square PSF shaping is employed. As in the original HoloTile publication \cite{madsen_holotile_2022}, square PSF shaping can be performed using an analytically computed phase profile, governed by $\beta$, a dimensionless parameter, given by:
\begin{equation}
	\beta = \frac{2\sqrt{2\pi}R \ell_{p-out}}{f\lambda}
\end{equation}
where $R$ is the $1/e^2$ beam waist radius of the incident light on the SLM.
Due to the wavelength dependence of the square beam shaping for any particular output pixel interval, a specific PSF shaping hologram is added to the tiled object hologram for each color channel to create the final color HoloTile holograms. 
This ensures that the width of the output pixels are identical to the output pixel interval $\ell_{p-out}$. The specific PSF shaping phases for each color channel, $\varphi_{psf}^r$, $\varphi_{psf}^g$, and $\varphi_{psf}^b$, in the case of $\ell_{p-out} = 30\text{ }\mu m$, are shown in \cref{fig:psf-shaping-rgb}. The differing phase curvature between the channels ensures that the effective pixel pitch is identical between the reconstructions.
Following calculation of both the tiled object holograms and the PSF shaping holograms, these are combined color-channel-wise, such that the result are three tiled and PSF shaped holograms; one for each illuminating wavelength.

\section{Experimental Results}
\begin{figure}[ht]
\begin{subfigure}[b]{.49\textwidth}
	\centering
	\includegraphics[width=.9\linewidth,page=1]{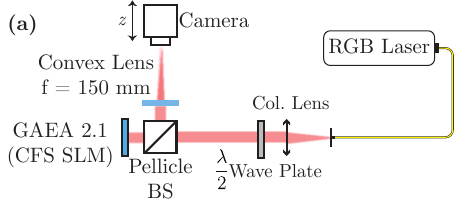}
	\captionsetup{width=.9\linewidth}
\end{subfigure}
\begin{subfigure}[b]{.49\textwidth}
	\centering
	\includegraphics[width=.9\linewidth,page=2]{Figures/optical-setup.pdf}
	\captionsetup{width=.9\linewidth}
\end{subfigure}
\captionsetup{width=.9\linewidth}
\caption{Experimental setup for capturing (a) \textit{lensed} and (b) \textit{lensless} HoloTile RGB reconstructions shown in this publication.}
\label{fig:optical-setup}
\end{figure}

\begin{figure}[ht]
	\centering
	\begin{subfigure}{0.585\textwidth}
	\includegraphics[width=\textwidth]{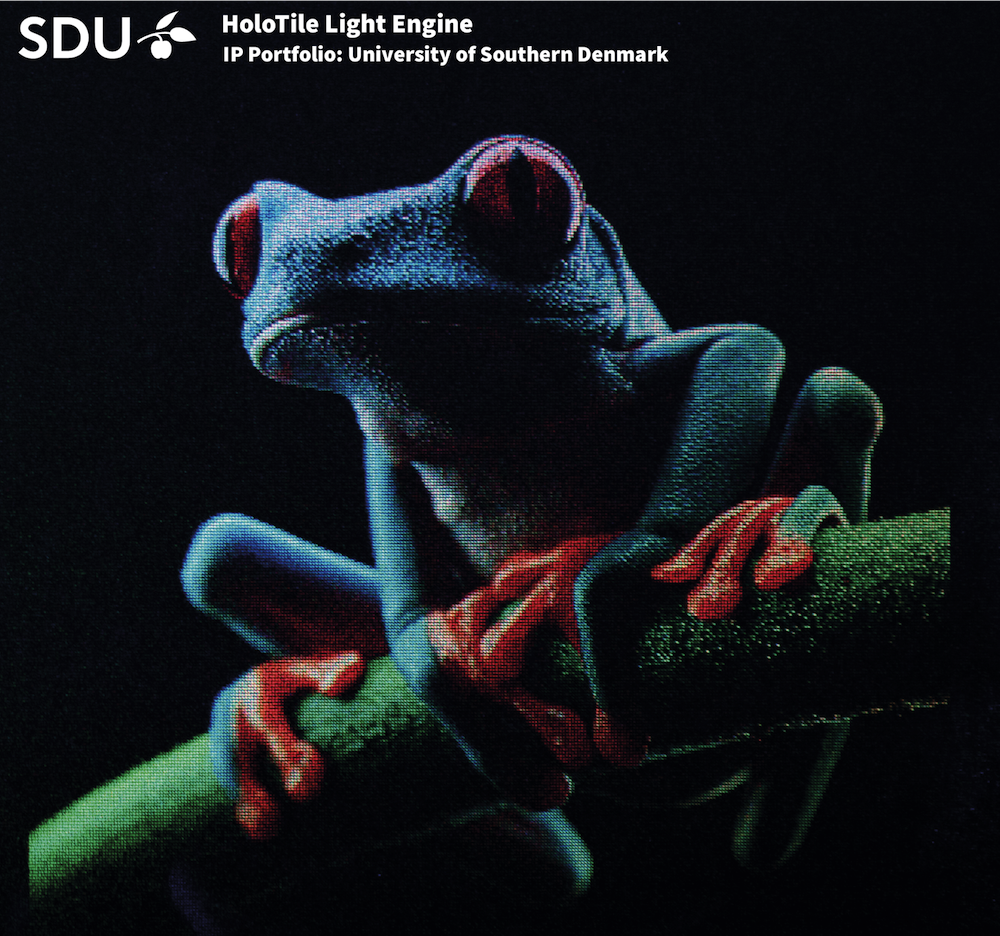}
	\end{subfigure}
	\begin{subfigure}{0.407\textwidth}
	\includegraphics[width=\textwidth]{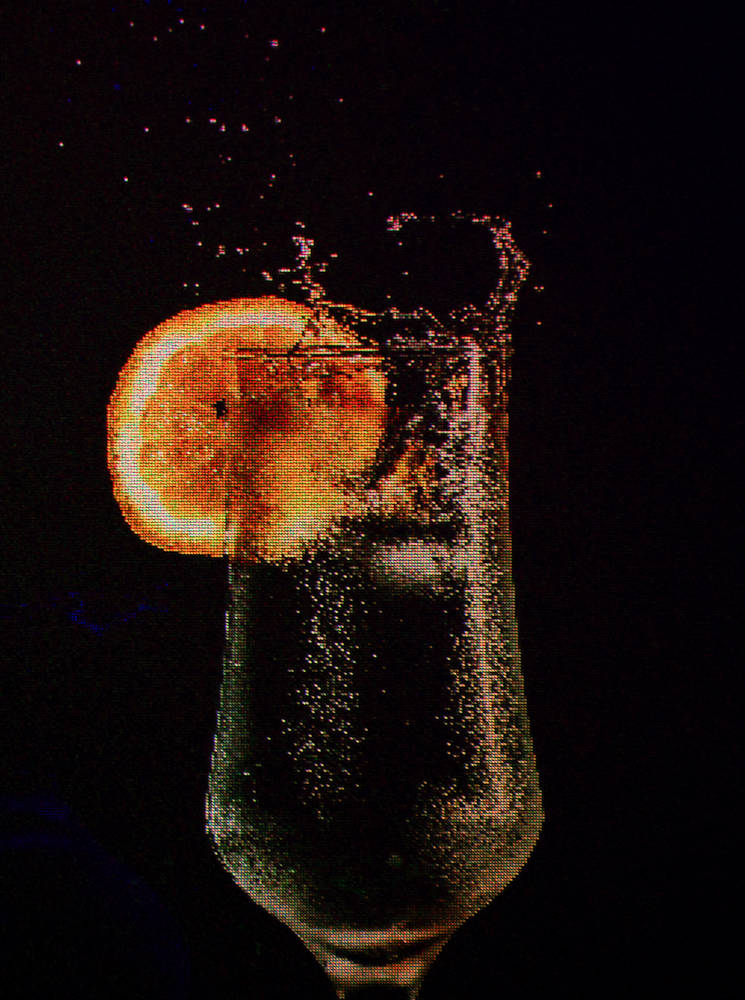}
	\end{subfigure}
	\par \smallskip
	\begin{subfigure}{0.495\textwidth}
	\includegraphics[width=\textwidth]{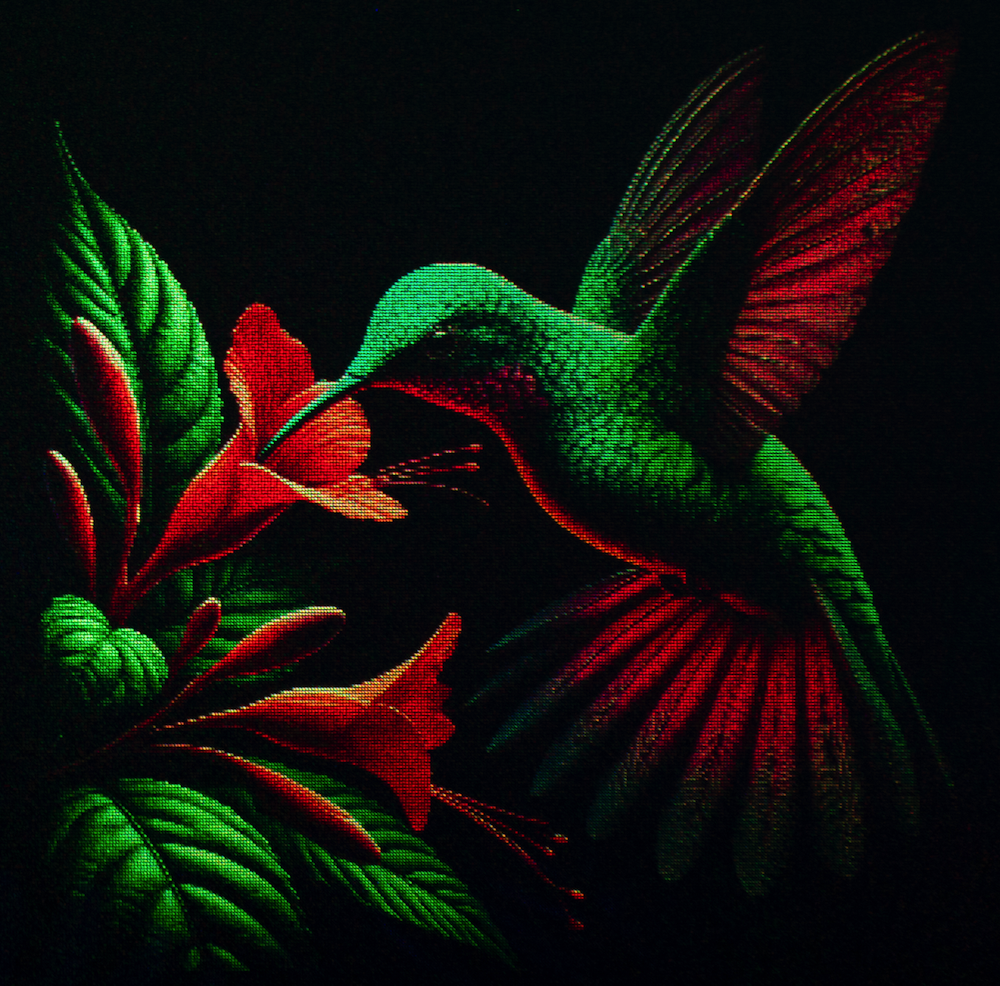}
	\end{subfigure}
	\begin{subfigure}{0.495\textwidth}
	\includegraphics[width=\textwidth]{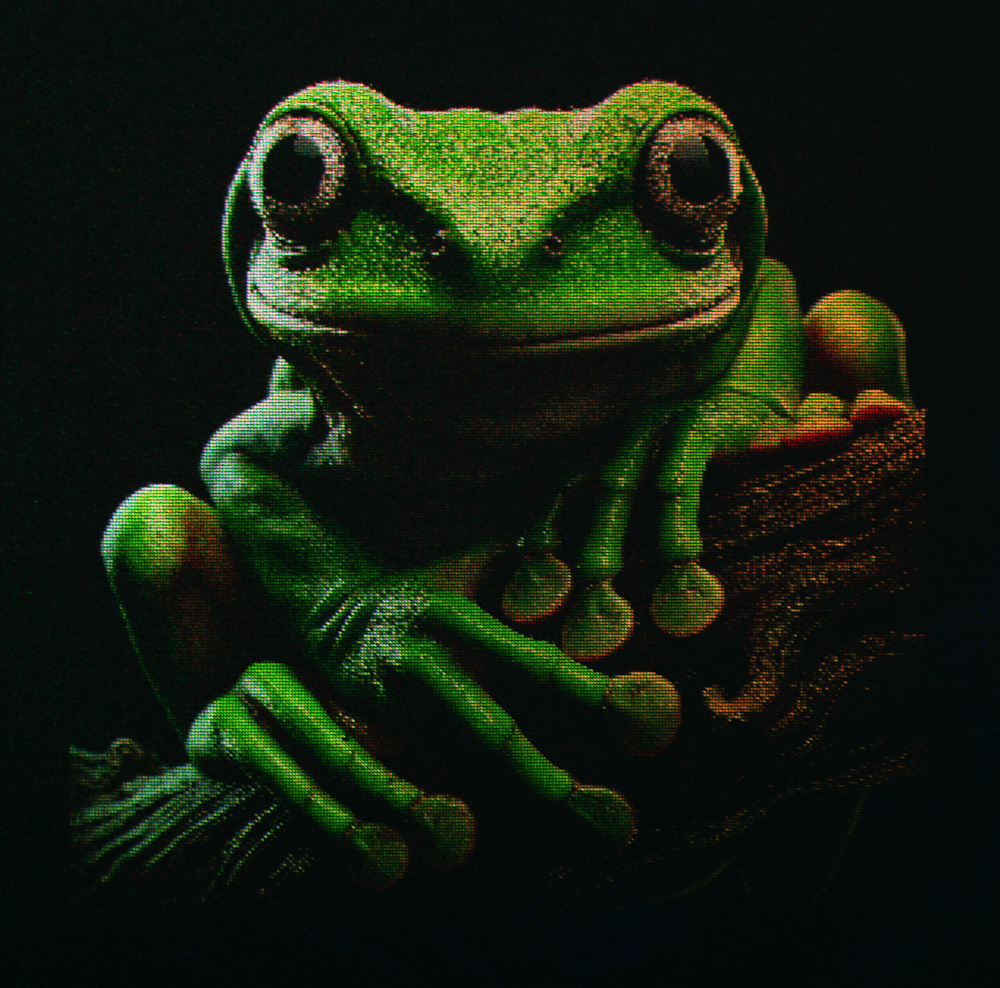}
	\end{subfigure}
	\captionsetup{width=.9\linewidth}
	\caption{Lensed experimental targets captured directly in camera in a single shot without temporal averaging. Holograms are added a linear phase to filter out the zero order.}
	\label{fig:lensed}
\end{figure}

The laboratory setup used to generate the results of this paper are shown in \cref{fig:optical-setup}. A FISBA ReadyBeam Ind2 RGB Laser (wavelengths: 450nm, 520nm, and 638nm) is collimated and shone on a HoloEye GAEA 2.1 SLM. The reflected read-out beam is directed by a pellicle beam splitter through a convex lens performing a spatial Fourier transform, the result of which is captured directly on a CMOS chip (Canon EOS M6 Mark ii).
Since the SLM is configured in CFS mode, each frame duration ($1/60s$) is divided into three subframes; one for each channel. The SLM is therefore cycling the color information at 180Hz.
Four captured reconstructions directly from the camera are shown in \cref{fig:lensed}. To remove the zero-order diffraction spot and isolate the first order, a linear phase ramp was added to each color channel in the RGB holograms. In the reconstructions, the pseudo-digital output pixels are clearly seen. Furthermore, the output pixels are well-aligned in terms of size and color channel scaling. In addition, high-frequency speckle noise, as commonly seen in both monochrome and RGB CGH, is also absent due to the PSF shaping.

\begin{figure*}[h]
	\centering
	\begin{subfigure}{0.652\textwidth}
	\includegraphics[width=\textwidth]{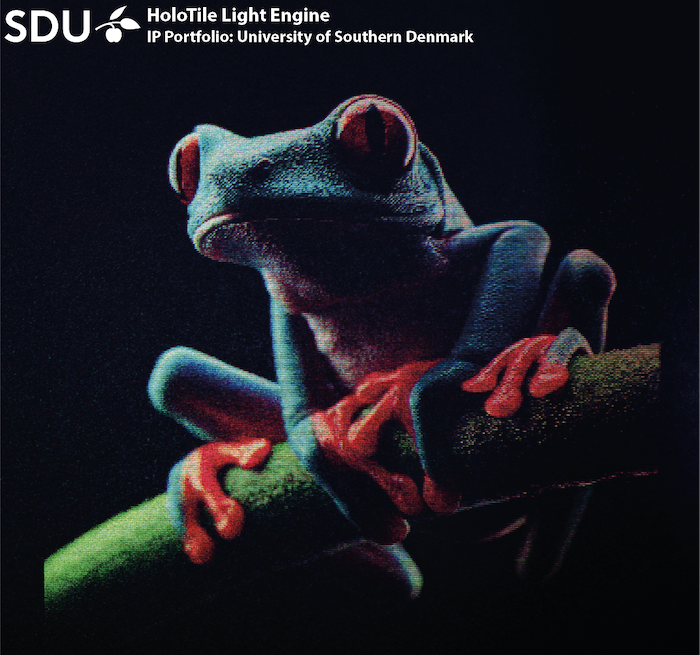}
	\end{subfigure}
	\begin{subfigure}{0.34\textwidth}
	\includegraphics[width=\textwidth]{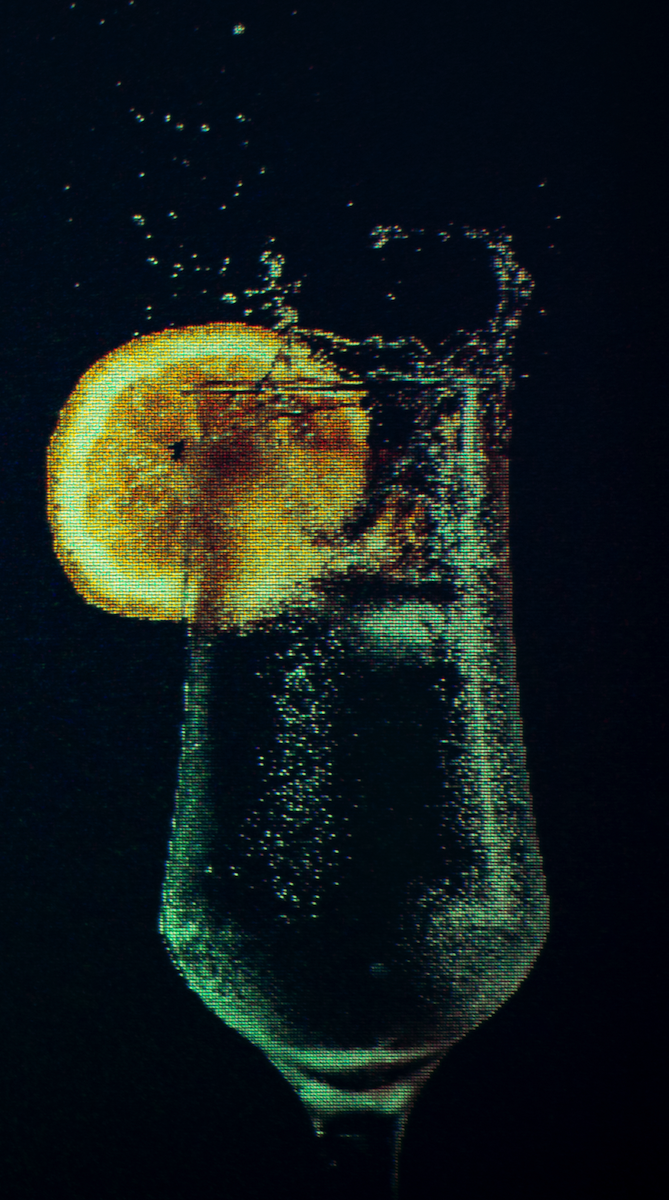}
	\end{subfigure}
	\par\smallskip
	\begin{subfigure}{0.489\textwidth}
	\includegraphics[width=\textwidth]{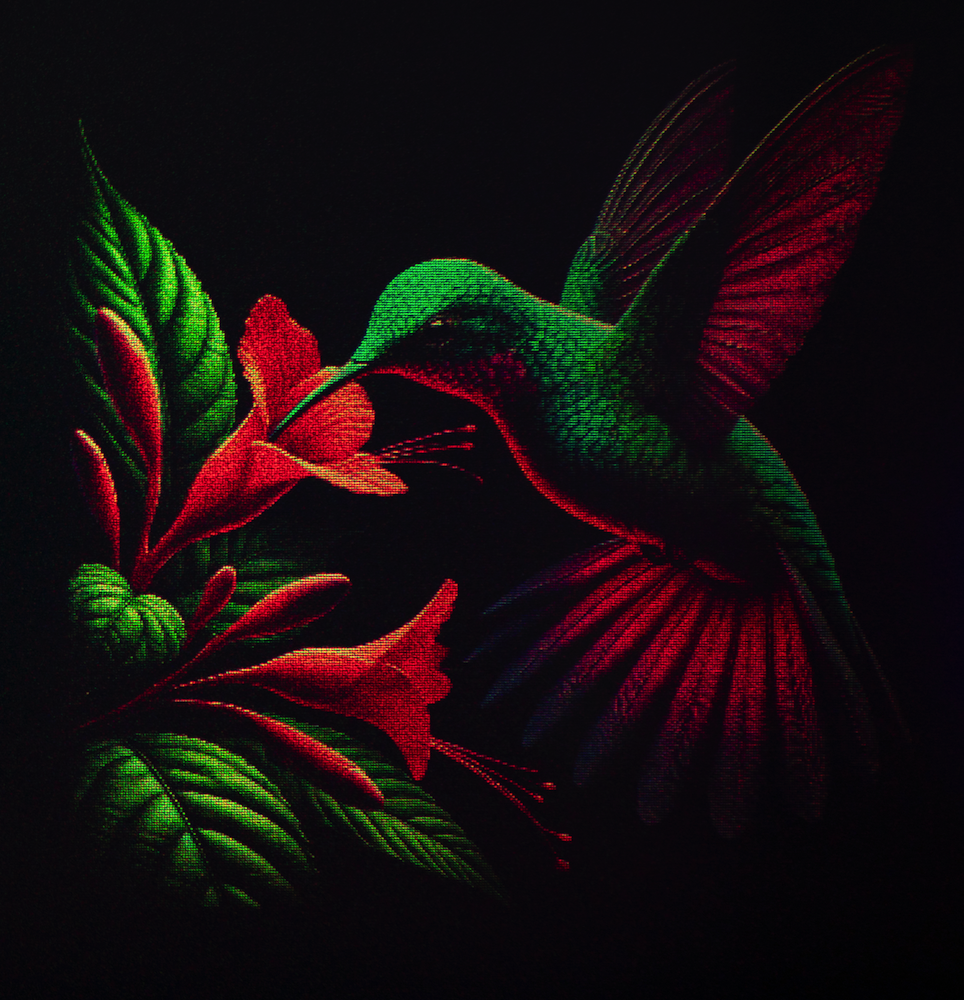}
	\end{subfigure}
	\begin{subfigure}{0.503\textwidth}
	\includegraphics[width=\textwidth]{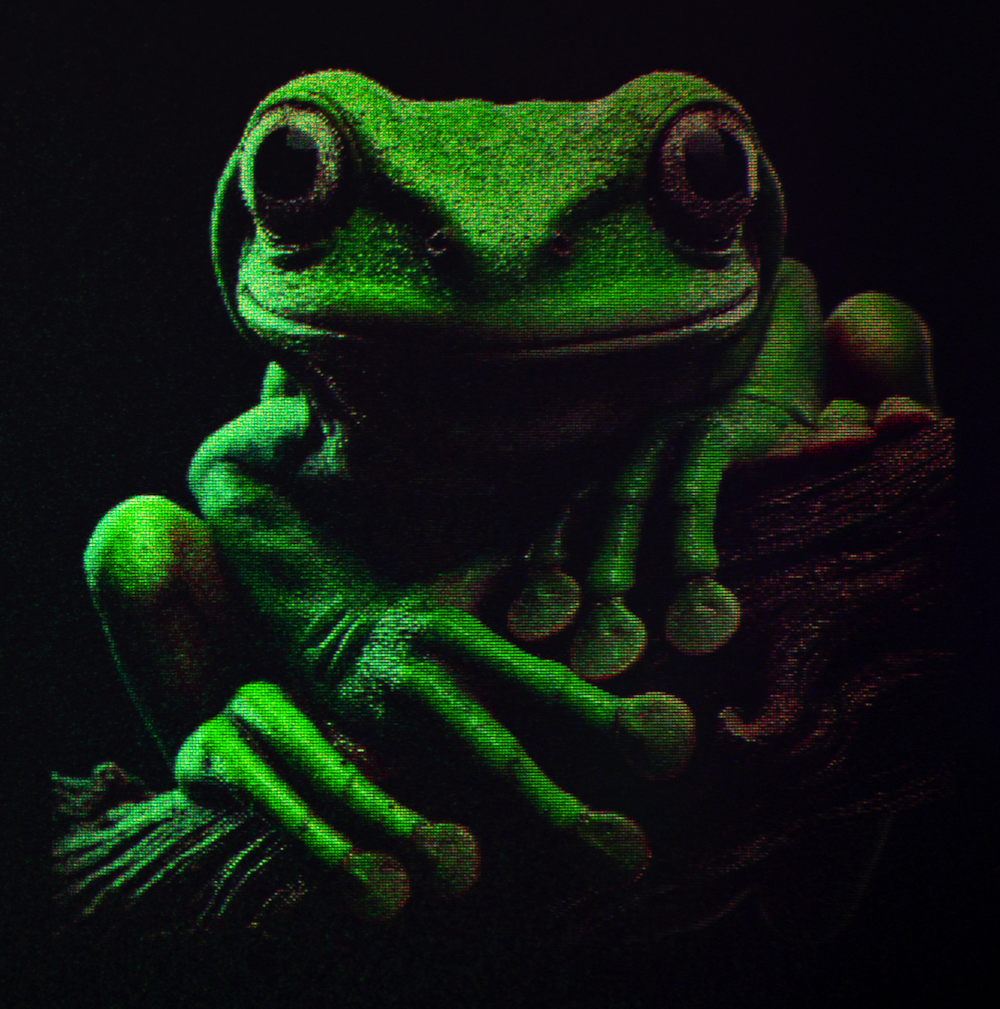}
	\end{subfigure}
	\captionsetup{width=.9\linewidth}
	\caption{Lensless experimental targets captured directly in camera in a single shot without temporal averaging. Holograms added a linear phase to filter out the zero order.}
	\label{fig:lensless}
\end{figure*}

\noindent
HoloTile can also function in a lensless configuration \cite{gluckstad_holotile_2024} by replacing the physical convex lens with an equivalent lens phase superimposed on the SLM (see \cref{fig:optical-setup}b). The lensless reconstructions as captured by the camera are shown in \cref{fig:lensless}. Here, the zeroth diffraction order is no longer a focused spot, but rather a large defocused white spot. Therefore, a linear phase ramp has been added to the lensless holograms as well in order to capture clean reconstructions. The lensless reconstructions also show the characteristic pseudo-digital output pixels, as well as good pixel alignment. However, due to white balance adjustments, color differences may be noticeable in \cref{fig:lensed,fig:lensless}.

\subsection{Color Video Reconstructions}
The reconstruction quality is not limited to still images. By repeating the hologram generation algorithm in sequence, the frames of a video can be used as hologram targets and displayed on the GAEA 2.1 SLM. Importantly, since the PSF shaping does not change over the course of the video, the PSF shaping hologram is only calculated for the first frame and is then re-used for the remaining video. With the lower resolution of the tiled subholograms, each frame can be calculated in synchronization with playback, thus allowing for real-time video display. 

In \cref{fig:frames}, select frames from the sequences in Video 1 in the Supplementary Material are shown. The video displays four sequences, ``Butterfly'', ``Fireworks'', ``Citrus'', and ``Nemo'', as they were captured by the recording camera (Canon EOS M6 Mark ii). Both ``Butterfly'' and ``Nemo'' display the ability of the approach to display full color information across the field-of-view of the scene, with very limited speckle. In ``Fireworks'' and ``Citrus'', the video contrast and detail level is highlighted.
\begin{figure}[ht]
\centering
\includegraphics{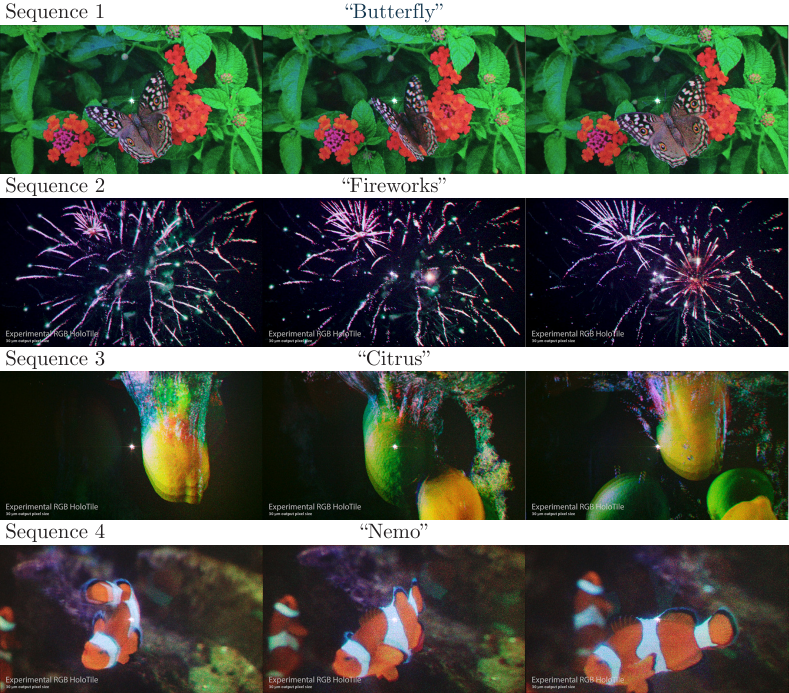}
\captionsetup{width=.9\linewidth}
\caption{Extracted experimentally captured video frames from the four sequences in Video 1 of the Supplementary Material.}
\label{fig:frames}
\end{figure}
In ``Citrus'', the fast moving fruits, along with the inherent difficulty of recording a video display source, causes a wavelength-dependent motion blur effect. However, this is not visible to the naked eye.
These results confirm that the proposed approach enables real-time holographic video playback while preserving spatial and color fidelity, with minimal speckle and stable phase reconstruction.

\section{Conclusion}
\noindent
In this paper, we introduced the first use of the HoloTile modality for multi-wavelength holography, demonstrating its capability to reconstruct high-fidelity, pseudo-digital RGB images with well-defined discrete output pixels. 
Leveraging subhologram tiling and point spread function (PSF) shaping eliminated the need for temporal averaging to suppress speckle, ensuring uniform output pixels across all wavelengths.

Our approach utilized a stochastic gradient descent (SGD) hologram generation algorithm for each wavelength, displayed sequentially on a HoloEye GAEA 2.1 Spatial Light Modulator. 
This setup achieved full 8-bit phase modulation at 60Hz for each wavelength, with reconstructions captured directly on a camera sensor, both for still images and video rate display.

The experimental results showcased the effectiveness of HoloTile in producing high-contrast reconstructions with superior homogeneity and regularity. The absence of high-frequency speckle noise, typically associated with conventional CGH, highlights the advantage of PSF shaping in our method. Additionally, the rapid hologram generation ($>$100x speed improvement \cite{gluckstad_holotile_2024}) positions HoloTile as a promising technique for real-time multi-wavelength applications in bio-photonics, additive manufacturing, and advanced display technology. While the experimental results shown in the previous sections are limited to RGB illumination for visual purposes, the multi-wavelength approach and process can be expanded to practically any combination of wavelengths, depending on the given application.

\section*{Acknowledgements}
\noindent
This work has been supported by the Novo Nordisk Foundation (Grand Challenge Program; 
NNF16OC0021948). 
We would like to express our gratitude to HoloEye Photonics AG for generously providing the HoloEye GAEA 2.1 Spatial Light Modulator used in this research.

\clearpage
\bibliography{references}

\end{document}